\documentstyle [11pt]{article}
\def\frac#1#2{{\textstyle{#1\over\vphantom2\smash{\raise.20ex
        \hbox{$\scriptstyle{#2}$}}}}}
\def\NP#1#2{Nucl. Phys. B\ {\bf{#1}},{#2}}
\def\PL#1#2{Phys. Lett. B\ {\bf{#1}},{#2} }

\def\PR#1#2{Phys. Rep. C\ {\bf{#1}},{#2} }
\def\PRL#1#2{Phys. Rev. Lett.\ {\bf{#1}},{#2}}
\def\PRD#1#2{Phys. Rev. D\ {\bf{#1}},{#2}}

\def\baselinestretch{1.5}
\def\title#1#2#3#4{
\begin{tabbing}
\= ~                 \hspace{4.5in} ~  \= COLO-HEP-
                                                  #1 \\
\>~                                  \> #2 \\
\end{tabbing}
\vspace{-0.6in}
\begin{center} {\large\bf #3}\\[.1in]
        {\bf #4}\\[.1in] {\it Physics Department, C.B. 390}\\
        {\it University of Colorado, Boulder, CO-80309}\\[.3in]
{\bf ABSTRACT}\\[-1.0in]
       \end{center}
\def\baselinestretch{1.1}\begin{quotation}}
\def\endtitle{\end{quotation} \newpage
}

\def\refer#1#2{{{#1}\markboth{REFERENCES}{REFERENCES}}
                \list{{\bf[}\arabic {enumi}{\bf]}}{\settowidth\labelwidth{[#2]}
                        \leftmargin\labelwidth\advance\leftmargin
                        \labelsep\usecounter{enumi}} 
                 }
\begin{document}
\hoffset=-1.0cm
\voffset=-0.5cm
\title
{271}
{~~}
{Differentiating various extra $Z^{\prime}$'s at Future Colliders.
}
{P. K. Mohapatra
\footnote{e-mail: pramoda@haggis.colorado.edu}
}
{We propose a way to differentiate between various extra $U(1)$ models
using flavor tagging in the decay modes $Z^{\prime}\rightarrow q{\bar{q}}$
once the extra $Z^{\prime}$ is observed at future colliders in the lepton
channel. A generalization of the
R parameter, namely, one for charge $\frac{1}{3}$ and one for charge
$\frac{2}{3}$ quark gives a two parameter test for the various models. Flavor
tagging eliminates the uncertainty because of extra fermions and can
reduce the QCD background at SSC/LHC dramatically.  For
$E_{6}$ and $SO(10)$ based models the former is  always 3. This seems to be a
very good way to eliminate certain models.\\
\vspace{1.3in}
\noindent
PACS No.s: 11.15, 12.10, 12.15
}
\endtitle
Although the standard model of electroweak interaction is in excellent
agreement
with the present experiments, there are reasons to believe in going beyond
the standard model. One of the ways to go beyond the standard model is to
extend
the gauge group.
Any extension of the standard model with a gauge group of rank bigger than four
entails one or more extra neutral vector bosons. The lightest of these is
generally known as $Z^{\prime}$ in the literature$^{1}$. There are two ways to
detect the presence of such a boson.
$Z^{\prime}$ through its mixing with the canonical $Z$
gives rise to various predictable deviations from the standard model results.
Most of the
effects of the extra neutral boson has been studied in great detail$^{1,2,3}$.
Many of these studies are exclusively concerned with the effects in high energy
experiments$^{2}$ and a few deal with atomic parity violation$^{3}$.
So far all the experimental observations agree very well
with the standard model predictions within the limits of experimental and
theoretical uncertainties. Hence, at the moment, one can only
put limits on the two parameters, the mixing angle of $Z^{\prime}$ with the
canonical $Z$ and the mass of $Z^{\prime}$. The experimental precision puts
a upper bound on the mass upto which the extra neutral gauge bosons effect
can be detected. These effects also depend on another parameter, namely
the mixing with the canonical $Z$. The limit is very much model dependent.
But it will be difficult to detect such an effect if the mass of the extra
neutral boson is more than 500 GeV. The other way is to look for direct
production and decay at future colliders. Such an extra $Z$ will be produced
at future $e^{+}e^{-}$ machines NLC/JLC and hadron colliders
SSC/LHC if its mass is less than 6 TeV$^{4}$. Once it is produced and detected
via its leptonic decays we need to know its origin. We need to differentiate
between the multitude of such models and possibly rule out many. There
have been lot of discussions on this subject. One crucial test is the
$A_{FB}$$^{5}$.
Recently there has been another proposal using rare decays$^{6}$ of
$Z^{\prime}$.
Because the number of extra $U(1)$ models available are really large it will
be helpful if we can find as many test parameters as we can. Here we propose
the generalization of the R parameter as a test for the various models.

Most of the studies on $Z^{\prime}$ decays and differentiating various
 $Z^{\prime}$ models have almost exclusively  been focussed on the lepton
sector. Very little attention has been paid to the decay mode
 $Z^{\prime}\rightarrow q{\bar{q}}$$^{7,8}$. The reason is the understandably
large (almost four orders of magnitude larger) QCD background in the hadron
colliders. No such disadvantage exist for the future generation $e^{+}e^{-}$
machines, NLC/JLC. Even in the hadron colliders the decay modes
$Z^{\prime}\rightarrow q{\bar{q}}$ deserve detailed analysis by
their own right. The  $Z^{\prime}$ study can't be complete by only
concentrating on the leptonic modes. The question is how to reduce the enormous
background ? Here we would like to give the proposal of flavor tagging.
Instead of looking for $Z^{\prime}\rightarrow jet+jet$ if we look for two
almost back to back jets which originated, let us say, from $b$ and $\bar{b}$
or from $t$ and $\bar{t}$ then we will have more handle on the background. The
techniques of identifying a $b$ jet by looking for high $p_{T}$ leptons already
exist and most probably will be improved before the hadron colliders produce
their first event. So, we propose to look for flavor tagging in the decay
$Z^{\prime}\rightarrow q{\bar{q}}$. This will give us better handle on reducing
the QCD background in the hadron colliders and will give two new parameters to
differentiate the origin of  $Z^{\prime}$. Both these points are discussed
in little more detail later on in this letter.

The neutral current interaction for any fermion
is given by\\
\begin{eqnarray}
{\cal{L}}_{NC}&=&-i \bar{\psi} \gamma^{\mu} \{eQA_{\mu} \nonumber \\
{}~ & ~ &-\frac{e}{\sin \theta_{W} \cos \theta_{W} }Z_{\mu}
         [(I_{3}-Q\sin^{2} \theta_{W}) \cos
           \chi +\frac{\cos \theta_{W} g_{2}}{g}X \sin \chi] \nonumber \\
{}~ & ~ &-\frac{e}{\sin \theta_{W} \cos \theta_{W} }Z^{\prime}_{\mu}
               [-(I_{3}-Q\sin^{2} \theta_{W}
           )\sin \chi + \frac{\cos \theta_{W} g_{2}}{g}X \cos \chi]\}\psi,
\end{eqnarray}
where $g$ and $g_{2}$ are the coupling constants for $SU(2)_{L}$ and the
extra $U(1)$ respectively; $X$ is the extra-$U(1)$ charge of $\psi$.

Here the X-charges of the
various fermions and the coupling constant of the extra $U(1)$, $g_{2}$, are
model dependent. For any given model the X-charges are exactly determined, upto
a constant multiplicative factor which can be absorbed into the coupling
constant. In models based on bigger unification group the coupling constant
$g_{2}$ is
related to $g$ and the relation depends on the details of the model such as
the original gauge structure and the representations of
Higgs bosons responsible for symmetry breaking.
We have tabulated the X-charges of the fifteen standard fermions in each family
for the eleven
models we have chosen for our analysis in Table 1. We can safely assume the
mixing angle to be very small and hence the interaction of the fermions with
$Z^{\prime}$ is given by\\
\begin{equation}
{\cal{L}}_{NC}~=~ i g_{2} X~\bar{\psi} \gamma^{\mu} Z^{\prime}_{\mu} \psi.
\end{equation}

The partial decay width of $Z^{\prime} \rightarrow \bar{f}f$ for any fermion
is given by
\begin{equation}
\Gamma ( Z^{\prime} \rightarrow \bar{f}f)~\propto ~C_{f}~(v_{f}^{2}+a_{f}^{2}),
\end{equation}
where $v_{f}$ and  $a_{f}$ are the vector and the axial vector couplings and
are respectively proportional to $X_{L}+X_{R}$ and  $X_{L}-X_{R}$ and $C_{f}$
is the color factor which is 3 for quarks and 1 for leptons. Although the
absolute branching fraction is uncertain by a a factor of 2 because of the
uncertainty of the spectrum of the exotic fermions in a model, the relative
branching fractions between the known fermions are very precisely known
theoretically for any particular model. So if we can distinguish between the
various exclusive decay modes$^{8}$
involving leptons and charge $\frac{1}{3}$ and charge $\frac{2}{3}$
quarks that will give us two very precise ratios. Flavor tagging is possible in
$e^{+}e^{-}$ machines and are relatively clean. The signature for $Z^{\prime}
\rightarrow jet+jet$
in the hadron colliders will be difficult
$^{7}$ because of the huge QCD background. The rough analysis shows that this
is in the borderline of being barely accessible. Our hope is that by flavor
tagging we can reduce the signal to background ratio dramatically.
A systematic study of this question has not been done yet.
The Monte-Carlo simulation of such events with appropriate cuts
and the relevant background using PYTHIA
are in progress$^{8}$. Preliminary results show that by putting appropriate
cuts on $p_{T}$ and invariant mass we have been able to reduce the
background to signal ratio from a number $\sim$ 10,000 to 40$\sim$50.

Here we assume that such identification is possible and
propose the following two
ratios of the branching fractions as a way to differentiate between the various
extra $U(1)$ models and possibly eliminate certain models.

\begin{equation}
R_{1}~=~{\Gamma {(Z^{\prime}\rightarrow \bar{b}b)}\over
        \Gamma ({Z^{\prime}\rightarrow l^{+}l^{-})}}
\end{equation}
and
\begin{equation}
R_{2}~=~{\Gamma {(Z^{\prime}\rightarrow \bar{c}c~or~\bar{t}t)}
   \over \Gamma ({Z^{\prime}\rightarrow l^{+}l^{-})}}
\end{equation}

We have analyzed eleven models for this report.
$S^{(0)}$ is the model recently proposed by Mahanthappa and the author$^{9}$
in which the X-charges are exactly proportional to the weak hypercharges(Y) for
the fifteen standard fermions. $S^{(i)}$ (i=1,2,3) come from $E_{6}$. They are
respectively called $\chi$, $I$ and $\eta$ in the literature ( see, for
example, the review by Hewett and Rizzo$^{1}$). $S^{(i)}$ (i=4,5) come from the
flipped $SU(5) \otimes U(1)$ broken to the standard model with the Higgs fields
residing in ($27+\bar{27}$)- and $78$- dimensional representations of $E_{6}$
respectively.  $S^{(6)}$ is the doubly flipped $SU(5) \otimes U(1) \otimes
U(1)$
with the Higgs fields residing
in ($27+\bar{27}$)-dimensional representations of $E_{6}$. $S^{(7)}$ has
its origin in the Pati-Salam group. $S^{(i)}$ (i=8,9,10) refer to the $SU(3)
\otimes U(1)$ models in which the extra third quark has electric charge
$\frac{2}{3}$, $\frac{1}{6}$ and $-\frac{1}{3}$ respectively$^{10}$. The models
$S^{i}$ (i=1..7) have been studied by Brahm and Hall$^{11}$ in connection with
dark matter. For the models $S^{(i)}$ (i=0..7) $g_{2}=0.8~c~g$ with the factor
$c$ being close to unity and for the models $S^{(i)}$ (i=8,9,10)
$g_{2}=\frac{g}{\cos \theta_{W}}$.

The $R_{1}$ and $R_{2}$ for the models discussed above are given in Table 2
and are plotted in Fig. 1. One remarkable fact that emerges is that $R_{1}$
is 3 for all the $E_{6}$ and thereby $SO(10)$ based models. So any model with
significant
departure from this value will be ruled out.
Combining this analysis with those involving $A_{F.B.}$ and rare decays
would further restrict the allowed models.

We thank G. Baranko and K.T. Mahanthappa  for
valuable comments, criticisms and discussions.
This work was supported in part by the Department of Energy under contract No.
DE-ACO2-86ER40253. \\
\newpage
\noindent
\refer{\bf References and Footnotes}{999}
\bibitem{rev} Some of the references are
             D. Iskandar and N.G. Deshpande, \PRL{19}{3457(1979)};
             U. Amaldi et. al., \PRD{36}{1385(1987)};
             V. Barger, N.G. Deshpande and K. Whisnant, \PRL{56}{30(1985)};
             V. Barger and K. Whisnant, \PRD{36}{979(1987)};
             R.W. Robinet and J.L. Rosner, \PRD{25}{3036(1982)};
             C.N. Leung and J.L. Rosner, \PRD{29}{2132(1982)};
             D. London and J.L. Rosner, \PRD{34}{1530(1986)};
             L.S. Durkin and P. Langacker, \PL{166}{436(1986)};
             G. Costa et. al., \NP{297}{244(1988)};
             J.L Hewett and T.G. Rizzo, \PR{183}{193(1989)}.
             The last article is a review.
\bibitem{lep} G. Altarelli et. al., CERN report no. 5752/90;
             J. Layssac, F.M. Renard and C. Verzegnassi, Lab. de Phy. Math.
             report no. LAPP-TH-290/90;
             K.T. Mahanthappa and P.K. Mohapatra, Proc. of the 25th.
             Int. Conf. on HEP, Aug. 1990(World. Scientific, Singapore) p. 966.
\bibitem{apv} K.T. Mahanthappa and P.K. Mohapatra, \PRD{43}{3093(1991)}, (E)
44,1616(1991);
              P. Langacker, \PL{256}{277(1991)}.
\bibitem{exs} J.L. Hewett and T. Rizzo, in {\it the proceedings of the 1988
Snowmass Summer study on HEP in the 1990's, Snowmass, CO 188};
              V. Barger et. al., \PRD{35}{166(1987)};
              F. Del Aguila, M. Quiros and F. Zwirner, \NP{287}{419(1987)};
              P. Chiappetta et. al., {\it Proc. LHC workshop, Achen, Germany,
              1990, pp-685}.
\bibitem{afb} P. Langacker, R. Robinett and J. Rosner, \PRD{30}{1470(1984)};
              J.L. Hewett and T.G. Rizzo, Madison Report No. MAD/PH/645.
\bibitem{rd}  M. Cvetic and P. Langacker, University of Pennsylvania Report No.
              UPR-487-T.
\bibitem{jpp} J. Pansart, Proceedings on the Workshop on LHC, Aachen,
4-9 Oct. 1990, ed. G. Jarlskog and D. Rein, CERN 90-10 vol. 2, p-709.
\bibitem{pkm} P.K. Mohapatra, to appear in the proceedings of Beyond the
Standard Model III, Ottawa, June22-24, 1992.
\bibitem{gb}  G. Baranko and P.K. Mohapatra, in preparation.
\bibitem{mm2} K.T. Mahanthappa and P.K. Mohapatra, \PRD{42}{1732(1990)}.
\bibitem{mm3} K.T. Mahanthappa and P.K. Mohapatra, \PRD{42}{2400(1990)}.
\bibitem{bh} D.E. Brahm and L.J. Hall, \PRD{41}{1067(1990)}.
\newpage
\begin{center}
{\bf Table 1}:
\small{X-charges of the fermions for the various extra $U(1)$ models. N is the
normalization factor.}\\
\vglue0.2in
\begin{tabular}{|c|c|c|c|c|c|c|c|}
\hline
Model  & $q_{L}$  & $u_{R}$ & $d_{R}$ & $l_{L}$ & $e_{R}$ & $\nu_{R}$ & 1/N \\
\hline
$S^{(0)}$ & 1 & 4 & -2 & -3 & -6 & 0 & 6 \\
\hline
$S^{(1)}$  & 1 & -1 & 3 & -3 & -1 & -5  & $\sqrt{40}$\\
\hline
$S^{(2)}$    & 2 & -2 & -4 & 4 & -2 & 0 &  $\sqrt{160}$ \\
\hline
$S^{(3)}$  & 4 & -4 & 2 & -2 & -4 & -10 &  $\sqrt{240}$ \\
\hline
$S^{(4)}$ & 0 & -2 & 0 & 2 & 2 & 0 &  4   \\
\hline
$S^{(5)}$  & 2 & 4 & -2 & -4 & -8 & -2 &  $\sqrt{96}$ \\
\hline
$S^{(6)}$  & 0 & -1 & 1 & 0 & 1 & -1 &  2 \\
\hline
$S^{(7)}$  & 1 & 1 & 1 & -3 & -3 & -3 &  $\sqrt{24}$ \\
\hline
$S^{(8)}$  & 0.29 &  0.11 & -0.05 &  0.19 & -0.16  &  0 & 1 \\
\hline
$S^{(9)}$  & 0.25 &  0  & 0 & 0.25 & 0 & 0 & 1 \\
\hline
$S^{(10)}$  & 0.24 &  -0.11 &   0.053 &  0.35 &  0.16 &  0 & 1 \\
\hline
\end{tabular}
\end{center}
\vglue1in
\begin{center}
{\bf Table 2:}
\small{Comparision of the branching ratios of the $Z^{\prime}$ in the
various models.}\\
\vglue.2in
\begin{tabular}{|c|c|c|}
\hline
Model & $r_{1}$  & $r_{2}$  \\
\hline
$S^{(0)}$ & 0.33  &  1.13       \\
\hline
$S^{(1)}$ & 3.00  &  0.60       \\
\hline
$S^{(2)}$ & 3.00  &  1.20        \\
\hline
$S^{(3)}$ & 3.00  &  4.80      \\
\hline
$S^{(4)}$ & 0.00  &  1.50      \\
\hline
$S^{(5)}$ & 0.30  &  0.75        \\
\hline
$S^{(6)}$ & 3.00  &  3.00       \\
\hline
$S^{(7)}$ & 0.30  &  0.30       \\
\hline
$S^{(8)}$ & 4.41  &  4.83       \\
\hline
$S^{(9)}$ & 3.00  &  3.00      \\
\hline
$S^{(10)}$ & 1.24  &  1.42      \\
\hline
\end{tabular}
\end{center}
\newpage
\vglue1in
\hglue1in
\begin{picture}(300,270)(0,0)
\put (0,0){\line(0,300){270}}
\put (0,0){\line(300,0){300}}
\put (143,-30){$R_{1}$}
\put (50,-1) {\line(0,100){2}}
\put (100,-1) {\line(0,100){2}}
\put (200,-1) {\line(0,100){2}}
\put (250,-1) {\line(0,100){2}}
\put (300,-1) {\line(0,100){2}}
\put (-1,50) {\line(100,0){2}}
\put (-1,100) {\line(100,0){2}}
\put (-1,150) {\line(100,0){2}}
\put (-1,200) {\line(100,0){2}}
\put (-1,250) {\line(100,0){2}}
\put (43,-10) {1.0}
\put (93,-10) {2.0}
\put (143,-10) {3.0}
\put (193,-10) {4.0}
\put (243,-10) {5.0}
\put (293,-10) {6.0}
\put (-40,143) {$R_{2}$}
\put (-20,43) {1.0}
\put (-20,93) {2.0}
\put (-20,143) {3.0}
\put (-20,193) {4.0}
\put (-20,243) {5.0}
\put (150,0){\line(0,300){270}} \put (155,260) {$E_{6}$}
\put (145.5,145.5){$\otimes$} \put (155,150) {(6,9)}
\put (145.5,25.5){$\otimes$}  \put (155,30) {(1)}
\put (145.5,55.5){$\otimes$}  \put (155,60) {(2)}
\put (145.5,235.5){$\otimes$} \put (155,240) {(3)}
\put (10.5,52.2){$\otimes$}   \put (20,60) {(0)}
\put (10.5,33.0){$\otimes$}   \put (20,40) {(5)}
\put (10.5,10.5){$\otimes$}   \put (20,15) {(7)}
\put (-4.5,70.5){$\otimes$}   \put (10,80) {(4)}
\put (216.0,237.0){$\otimes$} \put (225,245) {(8)}
\put (57.5,66.5){$\otimes$}   \put (65,75) {(10)}
\end{picture}\\
\vglue0.5in
\begin{center}
Fig. 1\\
$R_{1}$ and $R_{2}$ for the various models discussed in the text.
\end{center}
\end{document}